# An M dwarf accompanied by a close-in giant orbiter with SPECULOOS

Amaury H. M. J. Triaud,[1] Georgina Dransfield,[1] Taiki Kagetani,[2] Mathilde Timmermans,[3] Norio Narita,[4,5,6] Khalid Barkaoui,[3,6,7] Teruyuki Hirano,[5,8] Benjamin V. Rackham,[7,9]† Mayuko Mori,[2] Thomas Baycroft,[1] Zouhair Benkhaldoun,[10] Adam J. Burgasser,[11] Douglas A. Caldwell,[12,13] Karen A. Collins,[14] Yasmin T. Davis,[1] Laetitia Delrez,[3,15] Brice-Oliver Demory,[16] Elsa Ducrot,[17] Akihiko Fukui,[4,6] Clàudia Jano Muñoz,[18] Emmanuël Jehin,[15] Lionel J. García,[3] Mourad Ghachoui,[3,10] Michaël Gillon,[3] Yilen Gómez Maqueo Chew,[19] Matthew J. Hooton,[18] Masahiro Ikoma,[8] Kiyoe Kawauchi,[20] Takayuki Kotani,[5,8,21] Alan M. Levine,[9] Enric Pallé,[6] Peter P. Pedersen,[22] Francisco J. Pozuelos,[3,23] Didier Queloz,[18,22] Owen J. Scutt,[1] Sara Seager,[7] Daniel Sebastian,[1] Motohide Tamura,[5,8,24] Samantha Thompson,[18] Noriharu Watanabe,[2] Julien de Wit,[7] Joshua N. Winn[25] and Sebastián Zúñiga-Fernández[3]

*Affiliations are listed at the end of the paper*

## ABSTRACT

In the last decade, a dozen close-in giant planets have been discovered orbiting stars with spectral types ranging from M0 to M4, a mystery since known formation pathways do not predict the existence of such systems. Here, we confirm TOI-4860 b, a Jupiter-sized planet orbiting an M4.5 host, a star at the transition between fully and partially convective interiors. First identified with *TESS* data, we validate the transiting companion's planetary nature through multicolour photometry from the TRAPPIST-South/North, SPECULOOS, and MuSCAT3 facilities. Our analysis yields a radius of $0.76 \pm 0.02$ R$_{\rm Jup}$ for the planet, a mass of 0.34 M$_\odot$ for the star, and an orbital period of 1.52 d. Using the newly commissioned SPIRIT InGaAs camera at the SPECULOOS-South Observatory, we collect infrared photometry in *zYJ* that spans the time of secondary eclipse. These observations do not detect a secondary eclipse, placing an upper limit on the brightness of the companion. The planetary nature of the companion is further confirmed through high-resolution spectroscopy obtained with the IRD spectrograph at Subaru Telescope, from which we measure a mass of $0.67 \pm 0.14$ M$_{\rm Jup}$. Based on its overall density, TOI-4860 b appears to be rich in heavy elements, like its host star.

## 1 INTRODUCTION

Protoplanetary disc mass scales with stellar mass (e.g. Andrews et al. 2013). It is therefore expected that the lowest mass stars have less material available to form exoplanets with masses similar to Jupiter's. Many formation models confirm this (e.g. Laughlin, Bodenheimer & Adams 2004; Ida & Lin 2005; Alibert, Mordasini & Benz 2011; Burn et al. 2021). However, this does not mean that gas giant planet formation is totally impossible, and several examples have been identified with almost every detection method, including GJ 876 by radial velocities (Delfosse et al. 1998; Marcy et al. 1998; Correia et al. 2010), MOA 2003-BLG-53 by microlensing (Bond et al. 2004), and WASP-80 (Triaud et al. 2013) and HATS-6 (Hartman et al. 2015) by transits. The accumulation of these detections is being used to compute occurrence rates (e.g. Bryant, Bayliss & Van Eylen 2023) that suggest a small reduction in giant planet frequency compared to Sun-like stars, which will eventually feed into and calibrate population synthesis models (e.g. Burn et al. 2021) and improve our understanding of planetary formation and orbital evolution.

Of particular interest amongst all exoplanets are those in a transiting geometry. Wavelength-dependent transit observations obtained with high- and low-resolution spectra produce transmission spectra (Seager & Sasselov 2000). These are analysed to study exoplanets' atmospheric structure (Knutson et al. 2012), their chemical composition (Madhusudhan & Seager 2009), and even their wind speed (Snellen et al. 2010). Giant planets transiting M dwarfs are beneficial with respect to such planets transiting Sun-like stars in two main ways. First, the fractional amplitude of features within a planet's transmission spectrum scales with $R_\star^{-2}$, making a planet's atmospheric signature easier to measure. Secondly, M dwarfs' cooler effective temperatures allow short-period gas giants to have temperate equilibrium temperatures (e.g. Triaud 2021). Calibrating stellar evolution models for fully convective stars is important since the vast majority of exoplanets have physical parameters that depend on the assumed stellar parameters. The only practical way to obtain absolute stellar masses and radii is to model double-lined eclipsing binaries (e.g. Triaud et al. 2020). However, there are very few such systems among fully convective stars ($<0.3$ M$_\odot$; Baraffe et al. 2015). Such systems are intrinsically faint, short-period binaries are rare (Duquennoy & Mayor 1991), and the occurrence rate of binary stars decreases with the primary star's mass (Burgasser 2007). There is currently only one known double-lined, fully evolved eclipsing binary on the main sequence where the primary has a mass $<0.2$ M$_\odot$ (Casewell et al. 2018).

*E-mail: a.triaud@bham.ac.uk
† 51 Pegasi b Fellow.

To address these and closely related science topics, the SPECULOOS (Search for Planets EClipsing ULtracOOl Stars) consortium launched an internal project code-named MANGOs (M dwarfs Accompanied by close-iN Giant Orbiters with SPECULOOS; Dransfield, in preparation) whose purpose is to confirm and study all transiting/eclipsing companions to late M dwarfs. This confirmation of TOI-4860 b is the MANGOs project's first publication, for which we joined forces with the MuSCAT3 (Multicolor Simultaneous Camera for studying Atmospheres of Transiting exoplanets) and IRD (InfraRed Doppler) teams. TOI-4860 (2MASS J12141555-1310290, TIC 335590096, or *Gaia* DR2 3571038605366263424) is a $V = 16.47$, $J = 12.06$ M star at a distance of 80 pc (Stassun et al. 2019). This system contains a transiting/eclipsing object for which our analysis of all collected observations indicates a planetary companion. The system was first observed by *TESS* (Transiting Exoplanet Survey Satellite; Ricker et al. 2014) where a threshold-crossing event was detected. It is at that moment that TIC 335590096 received the TOI-4860 designation, with the planet candidate being referred to as TOI-4860.01. Several ground-based photometric observatories later confirmed the existence of periodic transit-like features. All photometric data are described in Section 2.1. In addition, spectra were collected to determine the stellar parameters and measure radial velocities (Section 2.2). We describe our analysis of all data in Section 2.3. Our results are detailed and discussed in Section 3.

## 2 OBSERVATIONS COLLECTED AND THEIR ANALYSIS

Another team has also confirmed this planetary system (Bonfils et al., submitted). Through communications with them, we agreed that they would report an analysis of both their data and the *TESS* data, and that this letter would report a fully independent analysis that does not cover *TESS* photometry. However, we do use *TESS* data to produce validation tests and search for transit timing variations (TTVs).

### 2.1 Description of the photometric data

TOI-4860 was observed with *TESS* in sector 10 with 30-min cadence spanning 2019 March 26 to 2019 April 22 and sectors 36 and 46 with 10-min cadence, respectively, spanning the periods 2021 March 7 to 2021 April 2 and 2021 December 2 to 2021 December 30. The full-frame images were first processed by the Quick Look Pipeline (Huang et al. 2020) resulting in the detection of a threshold-crossing event and the assignment of a *TESS* Object of Interest (TOI) number on 2021 December 21. The data were later reprocessed by the TESS-SPOC pipeline (Science Processing Operations Center; Jenkins et al. 2016) and we make use of these to search for TTVs. We retrieved these data from the NASA Mikulski Archive for Space Telescopes (MAST) with the LIGHTKURVE package (Lightkurve Collaboration 2018).

We obtained ground-based seeing-limited photometric data with the TRAPPIST-South/North (TRAnsiting Planets and PlanetesImals

Small Telescope; Gillon et al. 2011; Jehin et al. 2011; Barkaoui et al. 2019), SPECULOOS (Search for Planets EClipsing ULtracOOl Stars; Burdanov et al. 2018; Delrez et al. 2018; Sebastian et al. 2021), and MuSCAT3 (Multicolor Simultaneous Camera for studying Atmospheres of Transiting exoplanets; Narita et al. 2020) facilities. Two transits were obtained with TRAPPIST-South on 2022 March 27 and 2022 April 22 in a custom blue-blocking band and the *Sloan-z′* band, respectively, and we obtained one transit with TRAPPIST-North in the $I + z$ band on 2023 January 21. We observed one transit simultaneously in the *Sloan-g′*, *Sloan-i′*, and *Sloan-r′* bands, as well as using the SPeculoos' Infra-Red photometric Imager for Transits camera (SPIRIT), which is an InGaAs-based instrument, with a custom wide-pass filter called *zYJ* (Pedersen et al., in preparation) using all four SPECULOOS-South telescopes on 2023 January 27, and one transit in the *Sloan-z′* band on 2023 January 30. The data reduction and photometric measurements of the TRAPPIST and SPECULOOS data were done with a custom pipeline built with PROSE (Garcia et al. 2022), a PYTHON framework designed for image processing. MuSCAT3 observed one transit epoch on 2023 April 18 in four photometric bands simultaneously (*Sloan-g′*, *Sloan-i′*, *Sloan-r′*, and *Sloan-z$_s$*). All observations are reported in the journal of observations in Table A1 (Supplementary data).

### 2.2 Spectroscopic data and stellar parameter determination

We collected an optical spectrum of TOI-4860 with the MagE spectrograph (Magellan Echellette; Marshall et al. 2008) on the 6.5-m Magellan Baade Telescope on 2023 February 3 (UT) under clear and stable conditions.v We used the 0.7-arcsec slit to gather a 3500–10 000 Å spectrum with $R \sim 4100$. We collected two exposures of 530 s each, totalling 17.7 min on source, followed by a 2-s ThAr exposure. We observed the spectrophotometric standard LTT 3218 (Hamuy et al. 1994; Vernet et al. 2010; Moehler et al. 2014) earlier that night for flux calibration and did not apply any telluric correction. Relying on a standard set of afternoon MagE calibrations, we reduced the data with PYPEIT (Prochaska et al. 2020a, b). The final spectrum has a maximum Signal to Noise Ration (SNR) of 149 at 7584 Å and a median SNR of 61 over the full 3500–10 000 Å range.

The optical spectrum of TOI-4860 is best matched to M4 and M5 templates from Bochanski et al. (2007) (across the 4500–8900 Å range), so we adopt a mean classification of M4.5 ± 0.5. This classification is confirmed by spectral index/classification relations from Reid, Hawley & Gizis (1995), Gizis (1997), and Lépine, Rich & Shara (2003). The optical spectrum shows no indication of significant subsolar metallicity features, and measurement of the zeta metallicity index of Lépine et al. (2013), $\zeta = 1.113 \pm 0.003$, implies [Fe/H] = +0.15 ± 0.20 based on the calibration of Mann et al. (2013). None of the H I Balmer features are seen in emission or absorption, indicating weak magnetic activity, and we infer an upper limit of $\log_{10}(L_{H\alpha}/L_{bol}) \lesssim -5.4$ based on the $\chi$ relation of Douglas et al. (2014). The absence of magnetic emission suggests an age $\gtrsim 4.5$ Gyr (West et al. 2008). We also see no evidence of Li I absorption, consistent with an evolved low-mass star.

We collected a near-infrared spectrum of TOI-4860 with the SpeX spectrograph (Rayner et al. 2003) on the 3.2-m NASA Infrared Telescope Facility (IRTF) on 2022 April 19 (UT) in clear conditions with seeing of 1.5 arcsec. We used the short-wavelength cross-dispersed (SXD) mode with the 0.3 arcsec × 15 arcsec slit aligned to the parallactic angle, yielding spectra that cover 0.75–2.42 μm with $R \sim 2000$. We nodded in an ABBA pattern and collected 10 exposures of 179.8 s each, totalling 30 min on source. We collected the standard set of SXD flat-field and arc-lamp exposures immediately after the science frames, followed by an A0 V telluric standard. We reduced the data with SPEXTOOL v4.1 (Cushing, Vacca & Rayner 2004), following the standard approach. We use the SpeX Prism Library Analysis Toolkit (SPLAT; Burgasser & Splat Development Team 2017) to compare the spectrum to that of single-star spectral standards in the IRTF Spectral Library (Cushing, Rayner & Vacca 2005; Rayner, Cushing & Vacca 2009), finding the M5 standard Wolf 47 to be the best single match. The M4 standard Ross 47 only provides a marginally poorer fit, and so we adopt a spectral type of M4.5 ± 0.5, consistent with the optical classification. Following Delrez et al. (2022), we use index measurements across the SpeX spectrum defined by Mann et al. (2014) to derive a metallicity estimate of [Fe/H] = +0.45 ± 0.15.

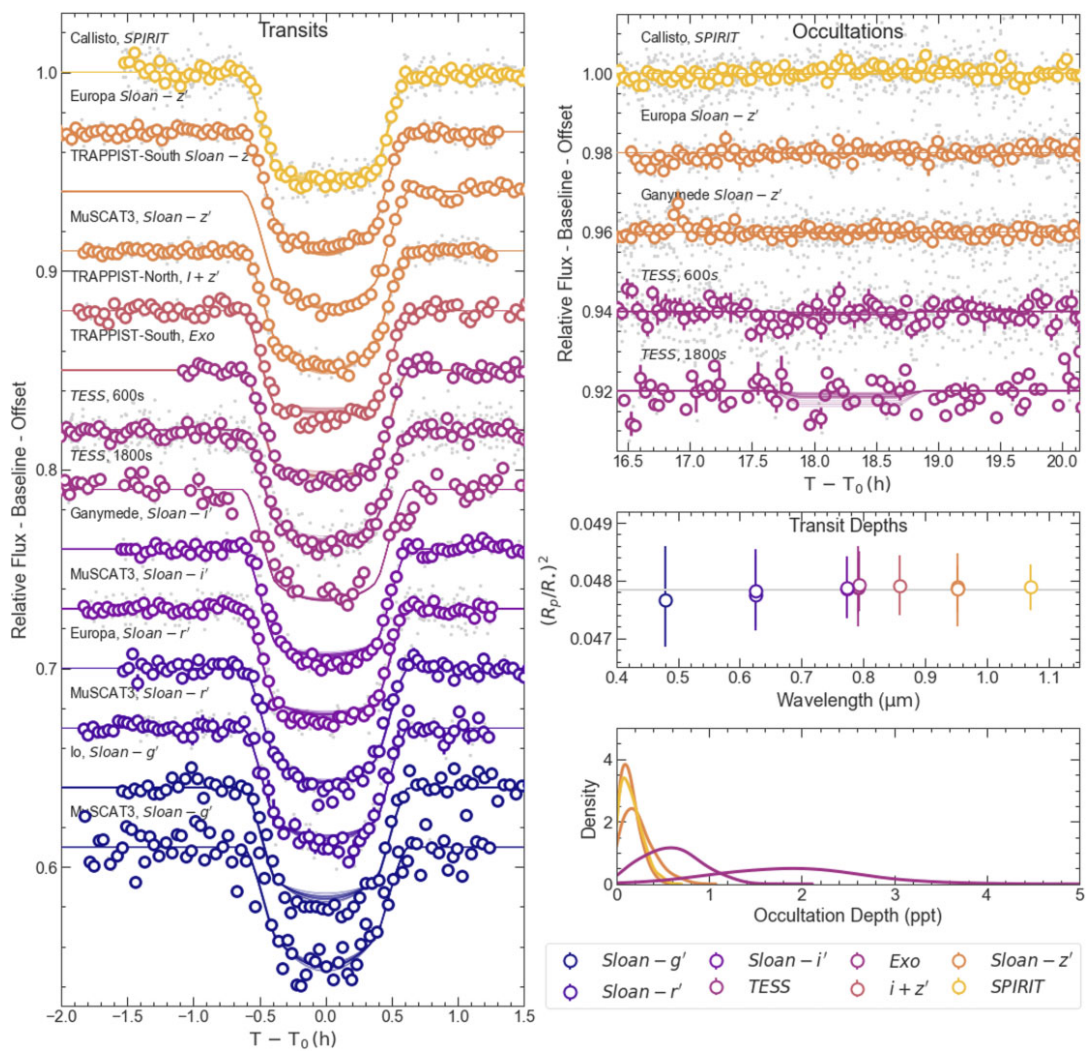

**Figure 1.** This figure shows all of the photometric data we use in this letter. On the left-hand side, we show each light curve and 20 models obtained from the posterior. On the top right, we plot photometry obtained around the time of secondary eclipse and 20 models from the posterior for each of the 14 light curves. The *TESS* data are not used to determine planetary parameters in this letter. Instead, they were used to corroborate the other observations we obtained. The middle right-hand panel shows that when transit depths $R_b^2/R_\star^2$ are fitted independently in each wavelength, they demonstrate that the transit signal is achromatic. The bottom right-hand panel draws the density of posterior samples for solutions including secondary eclipse depth parameters (arbitrary unit), assuming that the orbit of the planet is circular. All are consistent with no loss of light. All colours are chosen to depict the wavelength of the observations.

We use the infrared metallicity along with the *Gaia* Data Release 3 (DR3) distance and *BP–RP* colour (Gaia Collaboration 2016, 2023), and the 2MASS $K_s$-band apparent magnitude ($m_{K_s}$; Skrutskie et al. 2006) to derive the fundamental parameters of TOI-4860. With the *Gaia* distance and $m_{K_s}$, we calculate the absolute $K_s$-band magnitude ($M_{K_s}$). We then use the empirical relation between $M_{K_s}$, [Fe/H], stellar mass ($M_\star$; Mann et al. 2019, equation 5), and radius ($R_\star$; Mann et al. 2015, equation 5) to estimate $M_\star$ and $R_\star$. We estimate the stellar effective temperature ($T_{\rm eff}$) using the empirical relation between *Gaia BP–RP* colour, [Fe/H], and this parameter (Mann et al. 2015, equation 6). These and derivative parameters are given in Table A2 (Supplementary data).

In addition, seven high-resolution spectra were obtained with IRD (InfraRed Doppler; Tamura et al. 2012; Kotani et al. 2018), at Subaru, between 2023 January 8 and 2023 February 3. These data are reduced as described in Hirano et al. (2020) to obtain precision radial velocities with a mean precision of 23 m s$^{-1}$. Three of these were affected by poor weather, and two from technical issues, but we use all in our analysis and absorb their scatter with a jitter term. All measurements are listed in Table A3 (Supplementary data). Comparison of cross-correlation profiles of TOI-4860 with those of the reference star TOI-1634 (Hirano et al. 2021), also known to host a transiting planet (Cloutier et al. 2021; Hirano et al. 2021), show that TOI-4860 is a single-lined system (Fig. A3, Supplementary data). TOI-4860's line profile is similar to TOI-1634 implying a similarly slow rotating star. Based on the spectroscopic resolution of IRD, we can place a limit on the projected rotational velocity, $v \sin i_\star < 5$ km s$^{-1}$.

### 2.3 Joint analysis of all the data

We analyse the ground-based photometric and high-resolution radial velocity data described in Sections 2.1 and 2.2 using ALLESFITTER (Günther & Daylan 2019, 2021, and references therein), making use of the nested sampling algorithm DYNESTY (Speagle 2020).

We adopt the signal parameters reported by SPOC (Jenkins et al. 2016; Caldwell et al. 2020) as uniform priors and the stellar parameters derived in Section 2.2 as normal priors. We additionally make use of PYLDTK (Parviainen & Aigrain 2015) and the PHOENIX stellar atmosphere models (Husser et al. 2013) to calculate quadratic limb darkening coefficients for each photometric band used, and adopt these as normal priors.

We run two fits in the first instance: one with a circular orbit and the second where eccentricity is allowed to vary, parametrized as $\sqrt{e_b} \cos \omega_b$ and $\sqrt{e_b} \sin \omega_b$ (Triaud et al. 2011). We fit for all transit parameters ($R_p/R_\star$, $(R_\star + R_p)/a$, $\cos i$, $T_0$, and $P$) in both fits, as well as for the surface brightness ratio in the two photometric bands where we observed near phase 0.5, to search for the signal produced by a secondary eclipse. By doing so, we can derive an upper limit on the occultation depths and therefore on the brightness temperature of the planet. Such short orbits are likely to be circularized (e.g. Guillot et al. 1996), so we do not expect to find significant eccentricity in our model. We additionally fit for a radial velocity jitter term that is added in quadrature to the input RV errors.

Finally, to assess the presence of additional perturbing bodies in the system, we ran a third fit as in Dransfield et al. (2022), with each transit mid-time being a free parameter. We find that all transit mid-times measured from ground-based follow-up are consistent with no TTVs.

## 3 RESULTS AND DISCUSSION

Our joint analysis of the ground-based photometry and radial velocities confirms that candidate TOI-4860.01 is consistent with a transiting exoplanet. All transits we collected have similar depths, and the depths of odd transits are similar to those of even transits. All transits are consistent with the *TESS* transits as well (Fig. 1). To further confirm the planetary nature of this signal, we searched for contaminating light. *Gaia* measurements demonstrate that TOI-4860 is a high proper-motion system. A search of the Digitised Sky Survey data base[1] shows that there is no known contaminating background source near the target at the moment (Fig. A4, Supplementary data). In addition, mid-to-late M dwarfs suffer less from this kind of false positive on account of their spectral types. The IRD spectra also show that the system is single-lined, removing most gravitationally bound sources of contamination.

We find TOI-4860 b to be a 0.67-$M_{Jup}$, 0.76-$R_{Jup}$ planet on a $P_{orb} = 1.522$ d transiting orbit. The host star has an M4.5 spectral type ($T_{eff} = 3390$ K and $M_\star = 0.34$ M$_\odot$). As expected, measurements are compatible with a circular orbit, and we place a 95 per cent confidence upper limit at $e_b < 0.15$. All results are presented in Table 1, with complementary information in Table A4 (Supplementary data). These fitted and derived values were not obtained using any *TESS* data and are entirely independent of the parallel analysis by Bonfils et al. (submitted). The IRD radial velocities were mostly taken under adverse conditions, and we expect the mass to be weakly determined by our analysis (Fig. 2). TOI-4860 b's density of 1.55 $\rho_{Jup}$ implies that the planet is particularly enriched with heavy elements (e.g.

[1] Retrieved from https://irsa.ipac.caltech.edu/applications/finderchart on 2023 May 11.

**Table 1.** Fitted and derived parameters after performing a global analysis of the data. *TESS* photometry is not included in the analysis. The equilibrium temperature is calculated assuming an albedo of 0.3 and emissivity of 1.

| Fitted parameters | |
|---|---|
| $R_b/R_\star$ | $0.22009 \pm 0.00070$ |
| $(R_\star + R_b)/a_b$ | $0.10848 \pm 0.00085$ |
| $\cos i_b$ | $0.0197 \pm 0.0030$ |
| $T_{0,b}$ (BJD) | $2459832.641439 \pm 0.000032$ |
| $P_b$ (d) | $1.52275959^{+0.00000032}_{-0.00000035}$ |
| $K_b$ (km s$^{-1}$) | $0.250 \pm 0.049$ |
| $^a\sqrt{e_b} \cos \omega_b$ | $0.11^{+0.17}_{-0.24}$ |
| $^a\sqrt{e_b} \sin \omega_b$ | $0.056^{+0.10}_{-0.094}$ |
| $\ln \sigma_{jitter,ird}$ (ln km s$^{-1}$) | $-2.16^{+0.27}_{-0.25}$ |
| Derived parameters | |
| Host radius over semimajor axis, $R_\star/a_b$ | $0.08891 \pm 0.00066$ |
| Semimajor axis, $a_b$ (au) | $0.01845 \pm 0.00059$ |
| Inclination, $i_b$ (deg) | $88.87 \pm 0.18$ |
| Impact parameter, $b_{tra,b}$ | $0.221 \pm 0.033$ |
| Full-transit duration, $T_{full,b}$ (h) | $0.7742^{+0.0050}_{-0.0060}$ |
| Total transit duration, $T_{tot,b}$ (h) | $1.2435 \pm 0.0041$ |
| Epoch occultation, $T_{0,occ,b}$ | $2459833.402819 \pm 0.000033$ |
| $^a$Eccentricity, $e_b$ | $0.047^{+0.053}_{-0.035}$ |
| $^a$Argument of periastron, $w_b$ (deg) | $98^{+210}_{-82}$ |
| Companion mass, $M_b$ (M$_{Jup}$) | $0.67 \pm 0.14$ |
| Companion radius, $R_b$ (R$_{Jup}$) | $0.756 \pm 0.024$ |
| Companion density, $\rho_b$ (g cm$^{-3}$) | $1.92^{+0.50}_{-0.44}$ |
| Companion surface gravity, $g_b$ (cm s$^{-2}$) | $3090 \pm 610$ |
| Equilibrium temperature, $T_{eq,b}$ (K) | $624 \pm 30$ |

$^a$Parameters estimated by letting the eccentricity to be set by two free parameters. All other parameters assume a circular orbit.

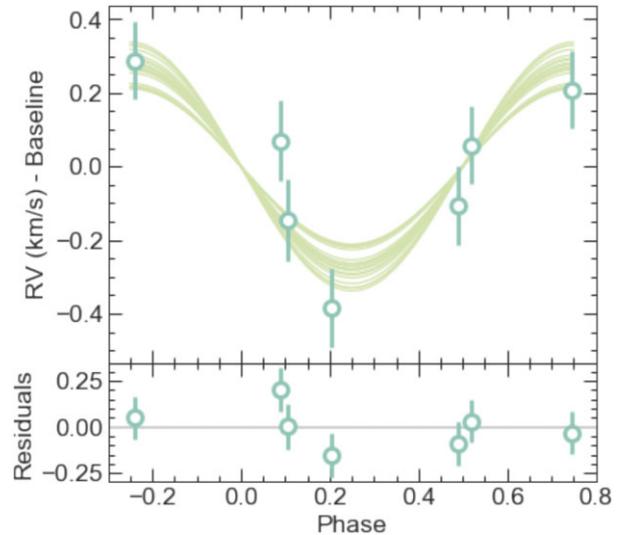

**Figure 2.** Radial velocities measured by IRD as a function of orbital phase. We show 20 models randomly drawn from the posterior.

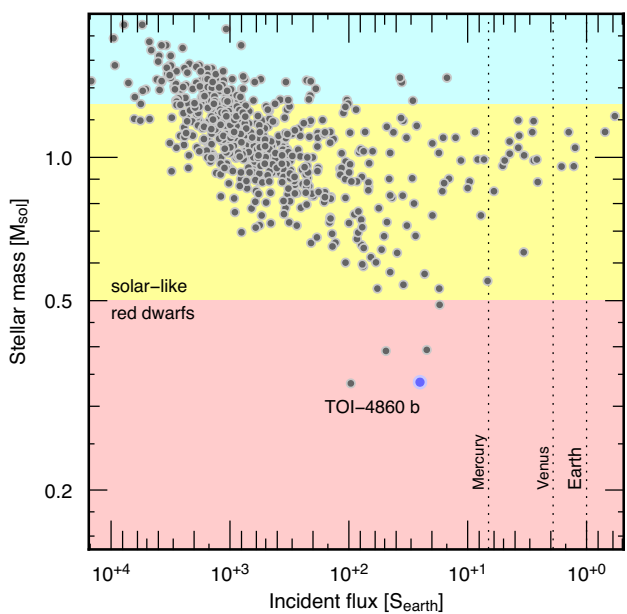

**Figure 3.** This diagram highlights TOI-4860 b's uniqueness amongst transiting giant exoplanets as being hosted by the lowest mass star yet. The purple dot corresponds to TOI-4860 b. All other grey dots correspond to planets with radius $>0.6$ $R_{Jup}$ and a measured mass $<25$ $M_{Jup}$. Data have been taken from TEPCAT (Southworth 2011).

Thorngren et al. 2016; Thorngren & Fortney 2019), which appears consistent with indications that the star is also enriched in metals (with a mean [Fe/H] = $+0.30 \pm 0.13$; Table A2, Supplementary data). Gas giants with large metal mass fractions are not rare, and TOI-4860 b likely falls into that category. Such systems are important to refine our understanding of the planet formation process (e.g. Morbidelli, Batygin & Lega 2023).

TOI-4860 b sits in a part of the parameter space that is very sparsely populated (Fig. 3). Restricting known transiting exoplanets to those where $R_p > 0.6$ $R_{Jup}$ and their host has $M_\star < 0.5$ $M_\odot$, there are seven objects listed on TEPCAT (Transiting ExoPlanets CATalogue; https://www.astro.keele.ac.uk/jkt/tepcat/; Southworth 2011). LHS 6343 (Johnson et al. 2011; Montet et al. 2015) and TOI-263 (Palle et al. 2021) both host a transiting brown dwarf. TOI-1227 b (Mann et al. 2022) does not have measured masses. TOI-1227 is an 11 Myr old system and its transiting object is suspected to have a mass in the range of Neptune's and still be contracting (Mann et al. 2022). Of the objects that remain, those most similar to TOI-4860 b are the very recently identified planets – TOI-519 b (Parviainen et al. 2021; Kagetani et al. 2023), TOI-3984 b (Cañas et al. 2023), TOI-3235 b (Hobson et al. 2023), and TOI-5205 b (Kanodia et al. 2023). All four orbit higher mass stars than TOI-4860, and receive similar irradiation. Thanks to its very short orbital period, TOI-4860 b provides a convenient opportunity to study the atmospheric properties of a warm Jupiter. We calculate the transmission spectroscopy metric (Kempton et al. 2018) to gauge how well this particular planet's atmosphere can be studied. Based on our fitted parameters, we find TSM = 67, a good score for this planet's equilibrium temperature.

Warm Jupiters have traditionally been defined as having orbital periods $>10$ d (e.g. Huang, Wu & Triaud 2016). This worked for Sun-like stars, but a more judicious definition would be to use the scaled distance $a_b/R_\star$ instead, which makes TOI-4860 b dynamically and irradiatively similar to other warm Jupiters. Warm Jupiters are expected to have followed a different migratory pathway from hot Jupiters, and a large fraction ($>50$ per cent) of them host additional planets in their system (Huang et al. 2016; Wu, Rice & Wang 2023). They also tend to occupy orbits that are aligned with the stellar equator (Triaud 2018; Albrecht, Dawson & Winn 2022; Rice et al. 2022; Attia et al. 2023).

**ACKNOWLEDGEMENTS**

We thank our reviewer and editor for a prompt and insightful review. Funding for the *TESS* mission is provided by NASA's Science Mission Directorate. We acknowledge the use of public *TESS* data from pipelines at the *TESS* Science Office and at the *TESS* Science Processing Operations Center. This research has made use of the Exoplanet Follow-up Observation Program website, which is operated by the California Institute of Technology, under contract with the NASA under the Exoplanet Exploration Program. This letter includes data collected by the *TESS* mission that are publicly available from the MAST. This research is based on data collected by the SPECULOOS-South Observatory at the ESO Paranal Observatory in Chile. ULiège's contribution to SPECULOOS received funding from the European Research Council under the European Union's Seventh Framework Programme (FP/2007-2013; grant agreement no. 336480/SPECULOOS), the Balzan Prize Foundation and the Francqui Foundation, the Belgian Scientific Research Foundation (F.R.S.-FNRS; grant no. T.0109.20), the University of Liège, and the ARC grant for Concerted Research Actions financed by the Wallonia-Brussels Federation. This work was supported by a grant from the Simons Foundation (PI: Queloz; grant no. 327127). This research was in part funded by the European Union's Horizon 2020 research and innovation programme (grant agreement no. 803193/BEBOP), and from the Science and Technology Facilities Council (STFC; grant nos. ST/S00193X/1 and ST/W000385/1). The material is based upon work supported by NASA under award number 80GSFC21M0002. The postdoctoral fellowship of KB is funded by F.R.S.-FNRS grant T.0109.20 and by the Francqui Foundation. BVR thanks the Heising-Simons Foundation for support. YGMC acknowledges support from UNAM PAPIIT-IG101321. FJP acknowledges financial support from the grant CEX2021-001131-S funded by MCIN/AEI/ 10.13039/501100011033. This publication benefits from the support of the French Community of Belgium in the context of the FRIA doctoral grant awarded to MT. LD is an F.R.S.-FNRS Postdoctoral Researcher. This research is based on data collected by the TRAPPIST-South telescope at the ESO La Silla Observatory. TRAPPIST is funded by the Belgian Fund for Scientific Research (Fond National de la Recherche Scientifique, FNRS) under the grant FRFC 2.5.594.09.F, with the participation of the Swiss National Science Fundation (SNF). This work was partly supported by MEXT/JSPS KAKENHI grant nos. JP15H02063, JP17H04574, JP18H05439, JP18H05442, JP19K14783, JP21H00035, JP21K13955, JP21K20376, and JP22000005, and JST CREST grant no. JPMJCR1761. This letter is based on observations made with the MuSCAT3 instrument, developed by the Astrobiology Center and under financial support by JSPS KAKENHI (JP18H05439) and JST PRESTO (JPMJPR1775), at Faulkes Telescope North on Maui, HI, operated by the Las Cumbres Observatory.

**DATA AVAILABILITY**

*TESS* data products are available via the MAST portal at https://mast.stsci.edu/portal/Mashup/Clients/Mast/Portal.html. Follow-up

photometry and high-resolution imaging data for TOI-4860 are available on ExoFOP (Exoplanet Follow-up Observations Programme) at https://exofop.ipac.caltech.edu/tess/target.php?id = 335590096. These data are freely accessible to ExoFOP members immediately and are publicly available following a 1-yr proprietary period. SPECULOOS and TRAPPIST data are also publicly available at the European Southern Observatory (ESO) archive a year after obtention.

The MagE Spectrograph User Manual is available at https://www.lco.cl/technical-documentation/the-mage-spectrograph-user-manual/. The SPEXTOOL User's Manual can be obtained from http://irtfweb.ifa.hawaii.edu/~spex/observer/.

## SUPPORTING INFORMATION

Supplementary data are available at *MNRASL* online.

**suppl_data**

Please note: Oxford University Press is not responsible for the content or functionality of any supporting materials supplied by the authors. Any queries (other than missing material) should be directed to the corresponding author for the article.

[1] School of Physics & Astronomy, University of Birmingham, Edgbaston, Birmingham B15 2TT, UK
[2] Department of Multi-Disciplinary Sciences, Graduate School of Arts and Sciences, The University of Tokyo, 3-8-1 Komaba, Meguro, Tokyo 153-8902, Japan
[3] Astrobiology Research Unit, Université de Liège, Allée du 6 août 19, B-4000 Liège (Sart-Tilman), Belgium
[4] Komaba Institute for Science, The University of Tokyo, 3-8-1 Komaba, Meguro, Tokyo 153-8902, Japan
[5] Astrobiology Center, 2-21-1 Osawa, Mitaka, Tokyo 181-8588, Japan
[6] Instituto de Astrofísica de Canarias (IAC), Calle Vía Láctea s/n, E-38200 La Laguna, Tenerife, Spain
[7] Department of Earth, Atmospheric and Planetary Sciences, MIT, 77 Massachusetts Avenue, Cambridge, MA 02139, USA
[8] National Astronomical Observatory of Japan, 2-21-1 Osawa, Mitaka, Tokyo 181-8588, Japan
[9] Department of Physics and Kavli Institute for Astrophysics and Space Research, Massachusetts Institute of Technology, Cambridge, MA 02139, USA
[10] Oukaimeden Observatory, High Energy Physics and Astrophysics Laboratory, Faculty of Sciences Semlalia, Cadi Ayyad University, Marrakech, Morocco
[11] Department of Astronomy and Astrophysics, UC San Diego, 9500 Gilman Drive, La Jolla, CA 92093, USA
[12] NASA Ames Research Center, Moffett Field, CA 94035, USA
[13] SETI Institute, 339 Bernardo Ave, Suite 200, Mountain View, CA 94043, USA
[14] Center for Astrophysics | Harvard & Smithsonian, 60 Garden Street, Cambridge, MA 02138, USA
[15] Space Sciences, Technologies and Astrophysics Research (STAR) Institute, Université de Liège, Allée du 6 Août 19C, B-4000 Liège, Belgium
[16] Center for Space and Habitability, University of Bern, Gesellschaftsstrasse 6, CH-3012 Bern, Switzerland
[17] AIM, CEA, CNRS, Université Paris-Saclay, Université de Paris, F-91191 Gif-sur-Yvette, France
[18] Cavendish Laboratory, JJ Thomson Avenue, Cambridge CB3 0HE, UK
[19] Instituto de Astronomía, Universidad Nacional Autónoma de México, Ciudad Universitaria, Ciudad de México 04510, México
[20] Department of Physical Sciences, Ritsumeikan University, Kusatsu, Shiga 525-8577, Japan
[21] Department of Astronomy, School of Science, The Graduate University for Advanced Studies (SOKENDAI), 2-21-1 Osawa, Mitaka, Tokyo 181-8588, Japan
[22] Department of Physics, ETH Zurich, Wolfgang-Pauli-Strasse 2, CH-8093 Zurich, Switzerland
[23] Instituto de Astrofísica de Andalucía (IAA-CSIC), Glorieta de la Astronomía s/n, E-18008 Granada, Spain
[24] Department of Astronomy, University of Tokyo, 7-3-1 Hongo, Bunkyo-ku, Tokyo 113-0033, Japan
[25] Department of Astrophysical Sciences, Princeton University, Princeton, NJ 08544, USA

This paper has been typeset from a T<sub>E</sub>X/L<sup>A</sup>T<sub>E</sub>X file prepared by the author.